\documentclass[conference]{IEEEtran}
\usepackage{ifpdf}
\usepackage{url}
\usepackage{amsmath}
\usepackage{subfigure}
\usepackage{amsmath}
\usepackage{comment}
\usepackage{color}
\usepackage{relsize}

\ifCLASSINFOpdf
  \usepackage[pdftex]{graphicx}
  \DeclareGraphicsExtensions{.pdf,.png}
\else
  \usepackage[dvips]{graphicx}
  \DeclareGraphicsExtensions{.eps}
\fi

\begin{document}

\title{Analysis of Human-Body Blockage in Urban Millimeter-Wave Cellular Communications}
\author{\thanks{This work is supported by Intel Corporation.}\IEEEauthorblockN{Margarita Gapeyenko$^{\dagger}$, Andrey Samuylov$^{\dagger}$, Mikhail Gerasimenko$^{\dagger}$, Dmitri Moltchanov$^{\dagger}$, Sarabjot Singh$^\star$,\\Ehsan Aryafar$^\star$, Shu-ping Yeh$^\star$,
Nageen Himayat$^\star$, Sergey Andreev$^{\dagger}$, and Yevgeni Koucheryavy$^{\dagger}$}
\IEEEauthorblockA{$^\dagger$W.I.N.T.E.R. Group, Tampere University of Technology, Tampere, Finland\\
$^\star$Intel Corporation, Santa Clara, CA, USA.
}}


\maketitle
\begin{abstract}
The use of extremely high frequency (EHF) or  millimeter-wave (mmWave) band has attracted significant attention for the next generation wireless access networks. As demonstrated by recent measurements, mmWave frequencies render themselves  quite sensitive to ``blocking" caused by obstacles like foliage, humans, vehicles, etc.  However, there is a dearth of analytical models  for characterizing such blocking and the consequent effect on the signal reliability. In this paper, we propose a novel, general, and tractable model for characterizing the blocking caused by humans (assuming them to be randomly located in the environment) to mmWave propagation as a function of  system parameters like transmitter-receiver locations and dimensions, as well as density and dimensions of humans. Moreover, the  proposed model is validated using a ray-launcher tool. Utilizing the proposed model, the blockage probability is shown to increase with human density and separation between the transmitter-receiver pair.  Furthermore, the developed analysis is shown to demonstrate the existence of a transmitter antenna height that maximizes the received signal strength, which in turn is a function of the  transmitter-receiver distance and their dimensions. 
\end{abstract}


\IEEEpeerreviewmaketitle

\section{Introduction and motivation}\label{sec:intro}


The dearth of spectrum in the conventional ultra-high frequency  (UHF) bands coupled with increasing wireless  traffic has led industry and academia to consider employing EHF or mmWave as one of the candidate technologies for the next generation of wireless access networks (or ``5G") \cite{Andrews2}. Given that electromagnetic waves cannot travel around obstacles with the dimensions exceeding their wavelength, numerous objects  (like humans, buildings, etc.) in the environment, which did not affect UHF signals significantly, would lead to propagation losses for mmWave transmissions \cite{propagation}. As high wireless traffic demand areas, attractive for deploying mmWave wireless networks, tend to be highly populated too (e.g., city square, mall, etc.), characterizing the effect of humans  around the receiver on the mmWave signal blockage is quite important.


Although there has been considerable progress in channel modeling for mmWave \cite{standard}, \cite{heath2}, \cite{SinJSAC14}, \cite{rappaport}, the investigation on blockage modeling for mmWave has been limited. In \cite{heath2}, a model was proposed to characterize the blocking of the line of sight (LoS) path by buildings. In the proposed model,  the receiver dimension  was assumed to be infinitesimally-small and the blockers (those high enough to block LoS) were distributed uniformly, without considering alternative deployment patterns. The derived model was quite similar to the exponential distance dependent decay approaches traditionally used in 3GPP urban outdoor micro-cellular model \cite{standard} and proposed in the context of mmWave in \cite{rappaport}. Another simple ball based blocking model for buildings was proposed in \cite{SinJSAC14}. Coverage and capacity obtained from these aforementioned models was compared with that obtained from real building data  in  \cite{andrews}.


Compared to UHF, communications in mmWave networks are expected to operate over shorter distances and in crowded urban environments \cite{rappaport}. Since the height of the transmitter could be much lower than that of  traditional base stations (BS), humans surrounding the receiver  can act as blockers to signal propagation. Hence, in addition to the heights of the receiver and the transmitter, the distance between them, and the spatial dimensions of blocking objects, the random heights of humans also need to be accounted for. Further, due to much smaller distances between the transmitter and the receiver, and accounting for the possible antenna arrays at the receiver \cite{rappaport}, the linear dimensions of the receiver may be non-negligible in practice. None of the prior works address these issues in their models. 



In this paper, we propose a novel and tractable  model for characterizing the probability of human-body blockage. The proposed model represents humans as cylinders with arbitrarily distributed heights and radii, whose centers follow the Poisson Point Process (PPP) in two dimensions. Employing the tools from stochastic geometry and renewal processes, the  blockage probability of the LoS path between transmitter-receiver is derived as a function of receiver dimension and the transmitter-receiver separation.  The case with infinitesimal receiver dimensions is a special scenario of the developed analysis. The proposed model is validated by comparing with  the blocking probability obtained from  detailed mmWave ray-launching simulations, which in turn lends credence to the offered design insights. Using our analysis, the optimal height of the mmWave transmitter in crowded outdoor environments is derived and shown to be proportional to the transmitter-receiver separation.



\begin{figure*}[t]
\centering
\subfigure[\label{fig:analytic_a}]
{\includegraphics[width=.3\textwidth]{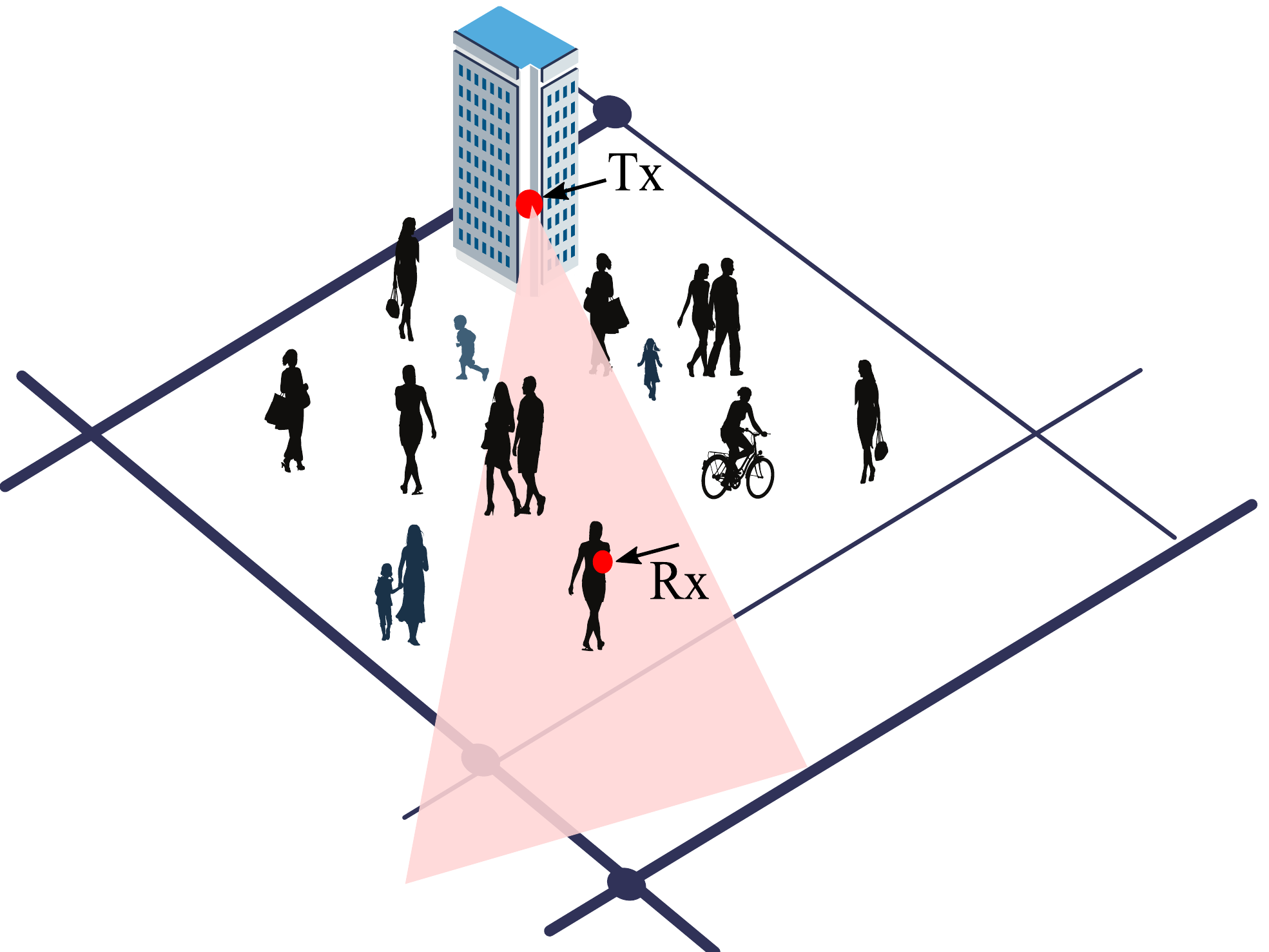}}
\subfigure[\label{fig:analytic_b}]
{\includegraphics[width=.3\textwidth]{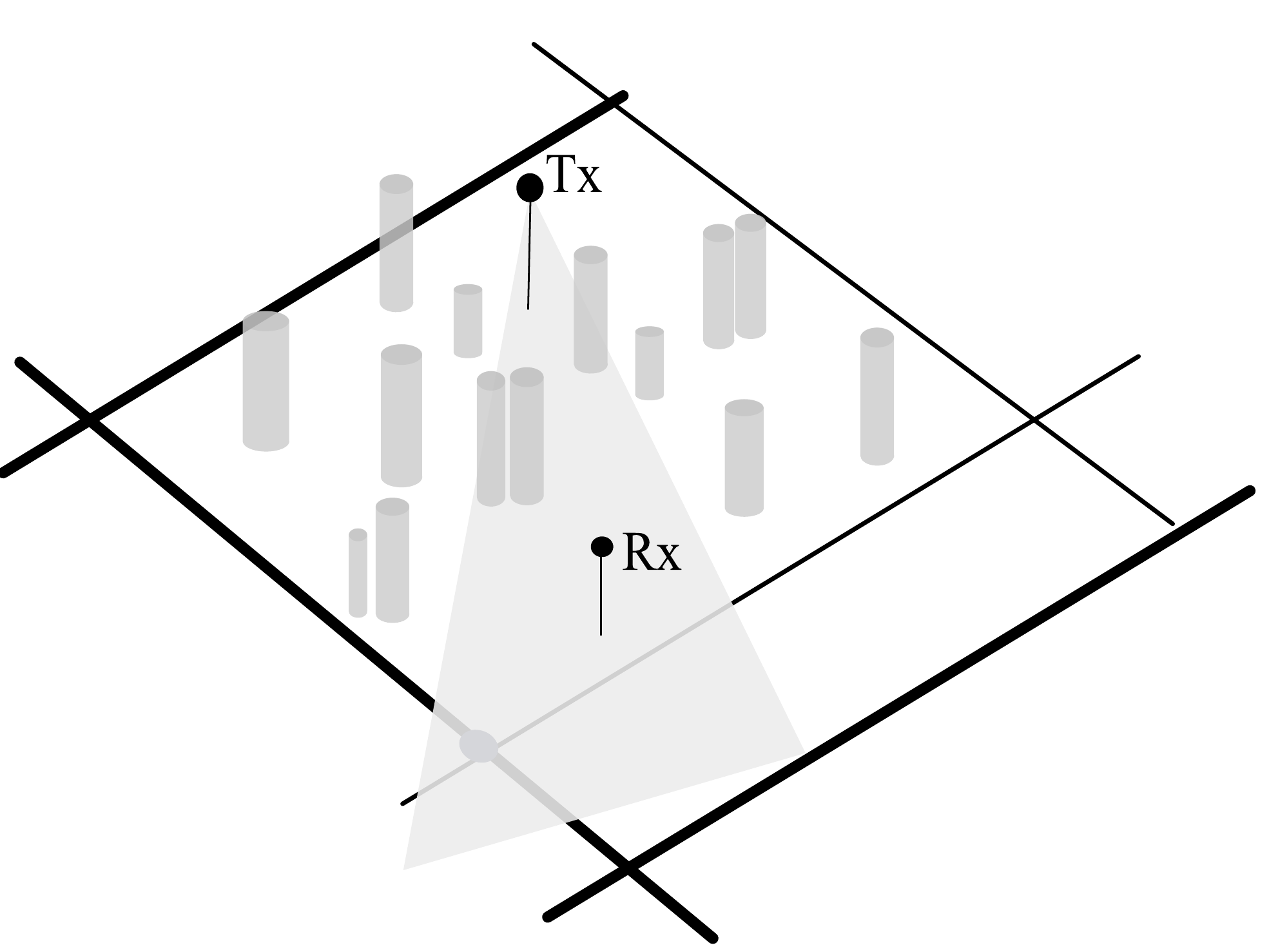}}
\subfigure[\label{fig:analytic_c}]
{\includegraphics[width=.3\textwidth]{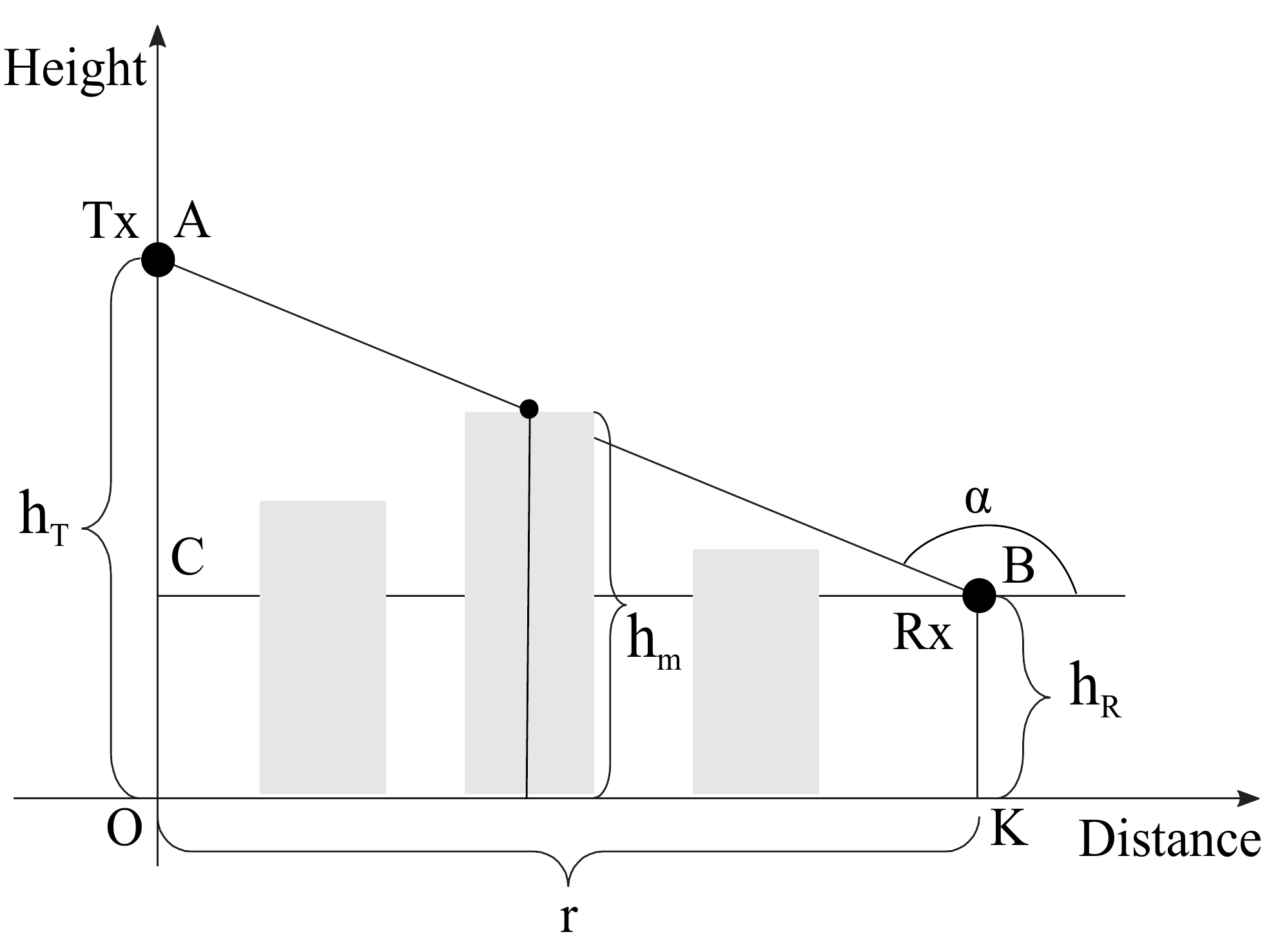}}
\caption{The considered scenario for analytical modeling.}
\label{fig:analytic}
\end{figure*}

\section{Proposed analytical framework}\label{sec:analysis}

\subsection{Spatial model}

Consider the scenario illustrated in Fig.~\ref{fig:analytic}(a, b). There is a Tx located at a certain height $h_{T}$ above the ground and a Rx located at the height $h_{R}$. The base of the Rx is at the distance $r$ from the base of the Tx. The potential blockers, humans, are distributed over the landscape. We model the blockers as cylinders \cite{cylinders} with a certain height, $H$, and the base diameter of $D$. Both $D$ and $H$ are random variables (RVs). It is known that the distribution of the height for men and women is Normal with the mean and the standard deviation provided in \cite{height}. Following \cite{height}, the mixture of users is closely approximated by the Normal distribution $H\sim N(\mu_{H}, \sigma_{H})$. Generally, any distribution could be used to provide a result based on the current methodology. The RV $D$ is assumed to be uniformly-distributed between $d_{min}$ and $d_{max}$. The centers of cylinder bases follow a Matern hard-core point process on the plane with the intensity $\lambda_{I}$. The length of the Rx is assumed to be $l$m. In summary, the main parameters and the description of the employed notation are given in Table \ref{tab:analysis_parameters}. In what follows, our main metric of interest is the probability of blockage for both the non-infinitesimal and infinitesimal receiver. 


\subsection{Blockage probability}

To represent the centers of blockers on the landscape, we employ the Matern hard-core process ensuring that the locations of blockers do not overlap. Using the results of \cite{Andrews1}, the Matern process can be replaced by the equivalent Poisson process for a wide range of intensities $\lambda_{I}$. Further, observe that for different values of $h_{T}$, $h_{R}$, and the distribution of the blocker heights $H$, not all the blockers affect the LoS between the Tx and Rx. The number of blockers should increase as the $x$-coordinate grows from $O$ to $r$, as in Fig.~\ref{fig:analytic_c}. The spatially-varying intensity of centers of blockers along the radial lines that may potentially affect the LoS between the Tx and Rx is given by
\begin{align}\label{eqn:001}
\lambda(x)=\lambda_{I}g(x),\,\,\,g(x)=Pr\{H>h_{m}(x)\}\,x\in(0,r),
\end{align}
where $h_{m}(x)$ is a function describing the distance between the line $AB$ and $OX$ at $x$.

Note that $h_{m}(x)$ is linear, $h_{m}(x)=ax+b$, where $a$ is the tangent of $h_{m}(x)$ with respect to the positive direction of $OX$, while $b$ is the height of a function at $x=0$. We thus have
\begin{align}\label{eqn:003}
h_{m}(x)=-\frac{h_{T}-h_{R}}{r}x+h_{T},\,\,\,x\in(0,r).
\end{align}

The probability $g(x)=Pr\{H>h_{m}(x)\}$ for each $x$ is a complementary cumulative distribution function (CCDF) of $H$. Since $H\sim{}N(\mu_{H},\sigma_{H})$, we have
\begin{align}\label{eqn:004}
g(x)=1-\frac{1}{2}\left[1+\text{erf}\left(\frac{h_{m}(x)-\mu_{H}}{\sigma_{H}\sqrt{2}}\right)\right],
\end{align}
where $\text{erf}(\cdot)$ is the error function.

\begin{figure}[!b]
\centering
\includegraphics[width=0.45\textwidth]{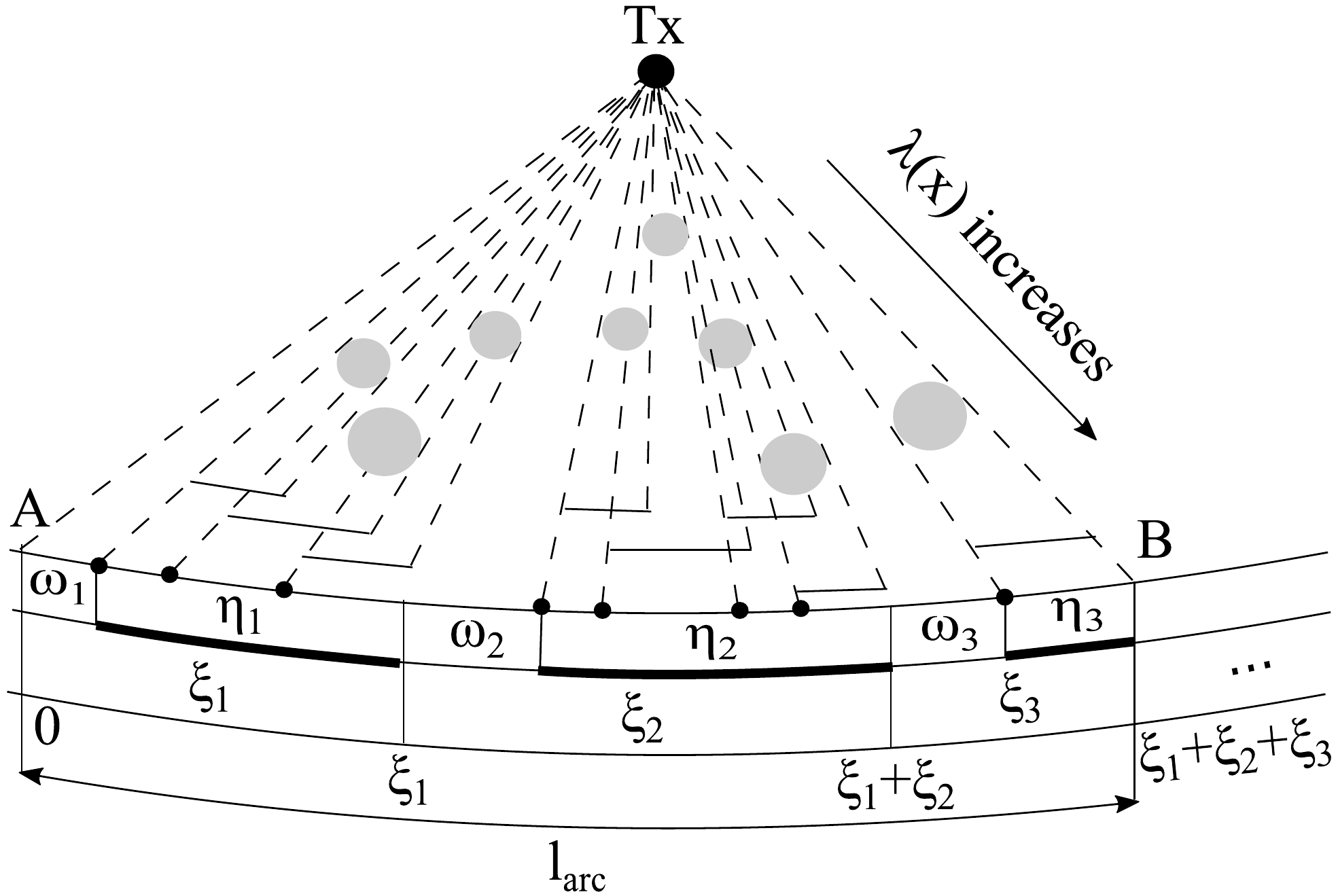}
\caption{Projections of blocker widths on the circumference.}
\label{fig:up}
\end{figure}

To determine the effective density of blockers at any separation distance $x$, the original homogeneous Poisson process is thinned with the probability $g(x)$. The resulting process is non-homogeneous, but still Poisson, with spatially-varying intensity along the radial lines, $\lambda(x)$ \cite{2}. The intensity $\lambda(x)$ is minimal at $x=0$ and increases non-linearly as $x$ grows. Consider now the projection of the blocker centers along the radial lines, represented by points on the circumference of the circle with radius $r$ and center at Tx, see Fig.~\ref{fig:up}. It is easy to prove that the process of projections on the circumference is homogeneous Poisson, as it has a Poisson distribution of projections in any bounded arc that depends only on the length of an arc and satisfies the independence property of the Poisson process. 


To establish the intensity of blocker centers at the circumference, consider the arc with length $l_{arc}$ as illustrated in Fig.~\ref{fig:up}. The mean number of points, $E[N_{B}]$, in the sector $ATxB$ is
\begin{align}\label{eqn:005}
E[N_{B}]= \int_{0}^{r}\lambda(x)x\frac{l_{arc}}{r}dx,
\end{align}
leading to the intensity of blockers at the circumference as
\begin{align}\label{eqn:006}
\mu=\frac{1}{l_{arc}}\int_{0}^{r}\lambda(x)x\frac{l_{arc}}{r}dx=\frac{\lambda_{I}}{r}\int_{0}^{r}xg(x)dx,
\end{align}
where $g(x)$ is given in (\ref{eqn:004}). Although this integral cannot be expressed in elementary functions due to the error function in $g(x)$, it can be easily computed numerically with any required accuracy.

\begin{figure}[!t]
\centering
\includegraphics[width=0.4\textwidth]{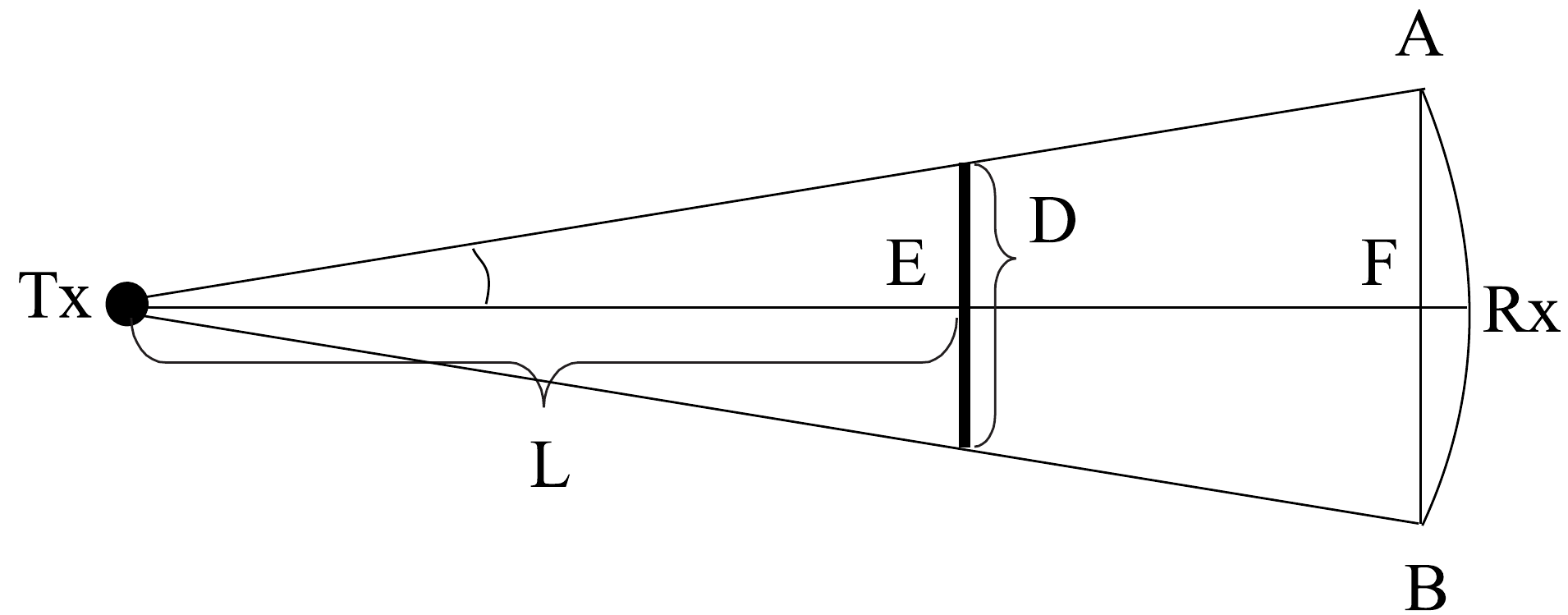}~~~~~
\caption{The top view of the scenario of interest.}
\label{fig:up1}
\end{figure}

To this end, we have characterized the point process of the centers of blockers. Further, the distribution of a "shadow" created by an individual blocker at circumference is given. Consider Fig.~\ref{fig:up1}, which shows the top view of our scenario. Observe that for $r>>D$, where $r$ is the distance from the base of Tx to Rx, we could replace the arc $ARxB$ by a chord $AFB$. From the geometric properties, we arrive at $W$, that is, a RV denoting the length of a shadow as
\begin{align}\label{eqn:007}
W=\frac{rD}{L},
\end{align}
where $L$ and $D$ are the RVs denoting the distance from the Tx to a blocker and  the width of a blocker, respectively. Recalling the principles of linear transformation of RVs \cite{ross}, the numerator of (\ref{eqn:007}) reads as
\begin{align}\label{eqn:008}
f_{rD}(x)=\frac{1}{r(d_{\max}-d_{\min})},\,\,\,x\in(rd_{\min},rd_{\max}).
\end{align}

Consider now the denominator of (\ref{eqn:007}). Recall that the intensity of blockers increases along the radial lines according to (\ref{eqn:001}). Therefore, the probability to have a blocker increases as we move from $x=0$ to $x=r$. The density to have a blocker at $x$ conditioned on the event that there is a blocker shall increase proportionally to $g(x)$, as obtained in (\ref{eqn:004}), and the only aspect we have to determine calculating the probability density function (pdf) of $L$ is the normalization constant, such that the area under $f_{L}(x)$ over $x\in(0,r)$ is exactly $1$. It can be found as
\begin{align}\label{eqn:009}
N=\int_{0}^{r}\left(1-\frac{1}{2}\left[1+\text{erf}\left(\frac{h_{m}(x)-\mu_{H}}{\sigma_{H}\sqrt{2}}\right)\right]\right)dx,
\end{align}
and normalized to obtain
\begin{align}\label{eqn:010}
f_{L}(x)=\frac{g(x)}{N},\,\,\,x\in(0,r).
\end{align}

Now, to determine the pdf of the shadow, we have to find the ratio between the RV $rD$ and $L$, whose densities are given by (\ref{eqn:008}) and (\ref{eqn:010}). Since these RVs are independent, the ratio is formally offered by \cite{ross}

\begin{align}\label{eqn:011}
f_{W}(y) =
 \begin{cases}
   \displaystyle\int_{\frac{rd_{min}}{y}}^{r}xf_{rD}(yx)f_{L}(x)dx, & \text{for}\\
   ~~~~~~~~~~                    \text{$d_{min}< y< d_{max}$}               & \\
   \displaystyle\int_{\frac{rd_{min}}{y}}^{\frac{rd_{max}}{y}}xf_{rD}(yx)f_{L}(x)dx, &\text{for}\\
   ~~~~~~~~~~~~~~~~~~~~\text{$y> d_{max}.$}&
 \end{cases}
\end{align}

The integral (\ref{eqn:011}) cannot be solved in elementary functions due to the density of $L$ in (\ref{eqn:010}). However, one can compute the distribution of $W$ numerically.

\subsection{Non-infinitesimal receiver}

Consider now the Rx of length $l$. In practice, it corresponds to when more than a single antenna is used at the user equipment and/or the distance between the bases of the Tx and Rx, $r$, is relatively small. In this case, we have to take into account the length of the Rx explicitly and the task at hand reduces to expressing the probability that an arc of a constant non-zero width $l$ is fully covered by arcs of random length, whose center points follow a Poisson process with the intensity $\mu$ as found in (\ref{eqn:005}) and with the length pdf $f_{W}(x)$ provided by (\ref{eqn:011}).

Consider the projections of blockers on the circumference as illustrated in Fig.~\ref{fig:up}. The widths of those projections are independent and identically distributed (i.i.d) RVs with the CDF $F_{W}(x)=\int_{-\infty}^{x}f_{W}(x)dx$ and the expected value $E[W]=\int_{-\infty}^{\infty}xf_{W}(x)dx$, where $f_{W}(x)$ is obtained from (\ref{eqn:011}). It is easy to show that not only the projections of the centers of blockers, but also their left- and right-hand side projections, form a stationary Poisson process on the line with the intensity $\mu$. The superposed process of all projections forms a renewal process with the alternating blocked and unblocked parts. An arbitrary point on the line is considered blocked, if it belongs to one of the blocked intervals. The question of blocking is then formulated as the probability of blocking this interval by the renewal process. An arc of length $l$ is said to be blocked, if all the points of this arc are blocked.

Let $\omega_{j}$ $\eta_{j}$, $j=1, 2, \ldots$ , denote the length of the unblocked and blocked intervals respectively, and define $\xi_{j}=\omega_{j}+\eta_{j}$. Points $0$, $\xi_{1}$, $\xi_{1} + \xi_{2}$, and $\xi_{1} + \xi_{2} + \xi_{3}$ are the renewal moments that form the renewal process. The density of this process is \cite{cox_Miller}, \cite{cox}
\begin{align}\label{eqn:51}
f(x)=\mu F_{W}(x)\exp\left(-\mu \int_{0}^{l}[1-F_{W}(y)]dy\right).
\end{align}

Let $f_{\xi}(t)$ be the density function of $\xi_{j}$, $j = 1,2, \ldots$ Functions $f_{\xi}(x)$ and $f(x)$ are related to each other via the renewal equation as \cite{cox_Miller}, \cite{cox}
\begin{align}\label{eqn:52}
f(x)=f_{\xi}(x) + \int_{0}^{l}f_{\xi}(x-y)f(y)dy.
\end{align}

The length of the unblocked part $\omega_{j}$ follows the exponential distribution with the parameter $\mu$, $F_{\omega}(x)=1-e^{-\mu x}$, with the mean $E[\omega]=1/\mu$ \cite{cox}. This can be verified by observing that the left-hand sides of the individual shadows follow a Poisson process with the intensity $\mu$. Hence, the distance from the end of the blocked part, considered as an arbitrary point, to the starting point of the next blocked interval is distributed exponentially. Let $F_{\eta}(x)$ and $F_{\xi}(x)$ be the CDFs of the length of the blocked intervals $\eta_{j}$, $j=1,2,\dots$, and the joint blocked/unblocked intervals, $\xi_{j}$, respectively, with the means $E[\eta]$ and $E[\xi]$. Further, let $F_{\eta}^{*}(s)$ and $F_{\xi}^{*}(s)$ be the corresponding Laplace-Stieltjes transforms (LSTs). For the joint interval $\xi_{j}$, we have
\begin{align}\label{eqn:53}
F_{\xi}^{*}(s)=F_{\eta}^{*}(s)F_{\omega}^{*}(s)=\mu \frac{F_{\eta}^{*}(s)}{\mu + s},
\end{align}
which can be solved for $F_{\eta}(x)$ in the RV domain as
\begin{align}\label{eqn:53}
F_{\eta}(x)=F_{\xi}(x)+\frac{f_{\xi}(x)}{\mu}.
\end{align}

Observe that the renewal density $f(x)$ is $f(x)=1/E[\xi]$, when $l\rightarrow \infty$. From (\ref{eqn:51}), we see that it is also equal to $f(x)=\mu\exp⁡(-\mu E[W])$, where $E[W]$ is the mean length of the blocked intervals. Consequently,
\begin{align}\label{eqn:54}
E[\xi]=\frac{1}{\mu}\exp(\mu E[W]).
\end{align}

Then, $E[\eta]$ can be established as
\begin{align}\label{eqn:155}
E[\eta]&=\int_{0}^{\infty}[1-F_{\eta}(x)]dx=\nonumber\\
&=\int_{0}^{\infty}\left(1-F_{\xi}(x)-\frac{f_{\xi}(x)}{\mu}\right)dx=E[\xi]-\frac{1}{\mu}.
\end{align}

\begin{table}[b!]\footnotesize
\centering
\caption{Description of notation and parameters}
\begin{tabular}{p{3.0cm}p{5.0cm}}
\hline
\textbf{Notation}&\textbf{Description} \\
\hline
$h_{T}$ & Height of Tx\\
$h_{R}$ & Height of Rx\\
$r$ & Distance between the bases of Tx and Rx\\
$l$ & Length of Rx\\
$H\sim~N(\mu_{H}, \sigma_{H})$ & Normally-distributed height of blockers\\
$D\sim~U(d_{min}, d_{max})$ & Uniformly-distributed width of blockers\\
$\lambda_{I}$ & Initial intensity of blockers\\
$\lambda(x)$ & Spatially-varying intensity of centers of blockers along the radial line\\
$g(x)$, $F_{H}(y)$ & CCDF, CDF of height of blockers\\
$\mu$ & Intensity of blockers at the circumference\\
$W$ & Length of a blocker's shadow\\
$f_{L}(x)$ & pdf of a distance between Tx and a blocker\\
$f_{W}(y)$, $F_{W}(y)$, $E[W]$ & pdf, CDF, and mean of a blocker's shadow\\
$\omega_{j}$, $F_{\omega}(x)$, $E[\omega]$ & Length, CDF, and mean of unblocked intervals\\
$\eta_{j}$, $F_{\eta}(x)$, $E[\eta]$  & Length, CDF, and mean of blocked intervals\\
$\xi_{j}$, $F_{\xi}(x)$, $E[\xi]$ & Length, CDF, and mean of $\omega_{j} + \eta_{j}$\\
$f(x)$ & pdf of renewal process\\
$f_{\xi}(x)$ & pdf of $\xi_{j}, j=1,2\ldots$\\
$f_{R}(y)$, $F_{R}(y)$ & pdf, CDF of blocker's radius\\
\hline
\end{tabular}
\label{tab:analysis_parameters}
\end{table}

Substituting (\ref{eqn:54}) into (\ref{eqn:155}), we arrive at
\begin{align}\label{eqn:55}
E[\eta]=\frac{1}{\mu}[\exp(\mu E[W])-1].
\end{align}
The probabilities that a random point on the line will be on the unblocked and blocked intervals, respectively, are the ratios of the corresponding parts, $E[\omega]/E[\xi]$ and $E[\eta]/E[\xi]$.

If a point is on the blocked part, then the distribution function of the length of the interval from this point to the right end of the blocked interval is
\begin{align}\label{eqn:56}
\tilde{F}_{\eta}(x)=\frac{1}{E[\eta]}\int_{0}^{l}[1-F_{\eta}(y)]dy.
\end{align}

Knowing the probability of blockage for a point, we can now obtain the probability of total blockage of the Rx of length $l$. Let $P(l)$ be the conditional probability that the interval $(0,l)$ is not blocked completely, meaning that the blocked interval containing the left-hand side of the Rx will end before $l$. The probability of the total blockage of the interval of length $l$, given that the left-hand side of this interval is blocked, can be found by using (\ref{eqn:53}), (\ref{eqn:55}), and (\ref{eqn:56}) as
\begin{align}\label{eqn:60}
&1- \tilde{F}_{\eta}(x)=\frac{1}{E[\eta]}\int_{l}^{\infty}[1-F_{\eta}(y)]dy =\nonumber\\
&=\frac{\mu}{\exp(\mu E[W])-1}\int_{l}^{\infty}\left(1-F_{\xi}(y)-\frac{1}{\mu}f_{\xi}(y)\right)dy,
\end{align}
and the probability that the left-hand side is blocked is $E[\eta]/E[\xi]$.

Finally, the probability of the total blockage follows from (\ref{eqn:60}) as
\begin{align}\label{eqn:61}
P_{B}&=1-P(l)=\mu \exp(-\mu E[W])\times{}\nonumber\\
&\times \int_{l}^{\infty}\left(1-F_{\xi}(y)-\frac{1}{\mu}f_{\xi}(y)\right)dy.
\end{align}

In the special case when the length of the receiver is less than the minimum diameter of the blocker, the probability of the total blockage is given by
\begin{align}\label{eqn:61a}
P_{B}&=1-\mu \exp(-\mu E[W])[1+\mu l].
\end{align}

\subsection{Infinitesimal receiver}

\begin{figure}[!ht]
\centering
\includegraphics[width=0.45\textwidth]{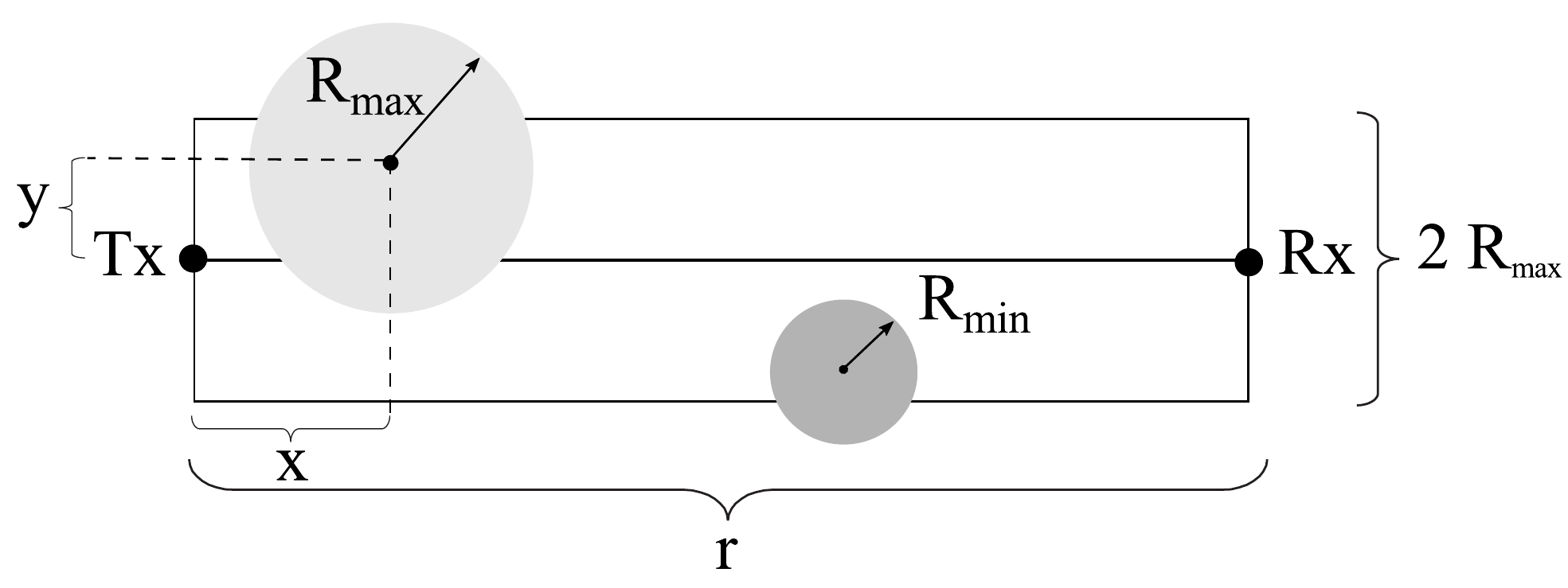}
\caption{Top view of the blocking area.}
\label{fig:fA}
\end{figure}

In many important cases, the size of the Rx can be assumed to be infinitesimally-small compared to other linear dimensions of objects. Equipped with this assumption, we provide a simpler method for calculating the LoS and blockage probabilities for a point receiver, when the width of the Rx is not considered. 
To this end, consider a rectangularly-shaped area as illustrated in Fig.~\ref{fig:fA}. Since this area is supposed to fit all the potential LoS blockers, its width is bounded by $d_{\max}$, that is, the maximum width of blockers. The length of the area is $r$. Due to the Poisson nature of blocker centers distribution, the number of blocker centers in the area of interest follows a Poisson distribution with the intensity $\lambda_{I}rd_{\max}$. Note that the coordinates of each particular center are uniformly-distributed over $(0,d_{\max})$ and $(0,r)$. To determine the probability of LoS blockage, we have to estimate the probability that at least one blocker, which is falling into the area of interest, blocks the LoS path.

Let the events $A_{i}$ define the probability of having $i$ blockers in the area of interest. Since each blocker has its own dimensions, for each of those we define the following events: (i) event $B_{0}$ that the radius of a blocker's base is not large enough to cross the LoS between the Tx and Rx and (ii) event $B_{1}$, which is complementary to event $B_{0}$. To calculate the probabilities of these events, we integrate over the blocker's $y$ coordinate as
\begin{align}\label{eqn:eA2}
Pr\{B_0\}=\int_{-r_{max}}^{r_{max}}f_R(y)F_R\left(\left|y\right|\right)dy,
\end{align}
and $Pr\{B_{1}\}=1-Pr\{B_{0}\}$, where $f_R(y)$ and $F_{R}(y)$ are the pdf and CDF of the blocker's radius, respectively. The radius of a blocker is uniformly-distributed in $(d_{\min}/2,d_{\max}/2)$.

Recall that the radius and height of a blocker are independent. Define the following events: (i) event $C_{0}$ that the blocker is not high enough to block the LoS and (ii) event $C_1$, which is complementary to event $C_{0}$.  The probabilities of these events are
\begin{align}\label{eqn:eA4}
Pr\{C_0\}=\int_{0}^{r}f_R(x) F_H\left(\frac{h_{T}r-(h_{T} - h_{R})x}{r}\right)dx,
\end{align}
and $Pr\{C_1\}=1-Pr\{C_0\}$, where $f_R(x)$ is the pdf of a uniform distribution from $0$ to $r$, and $F_H(y)$ is the CDF of the blocker's height. 

Having defined all the events of interest, we can proceed with obtaining the probability of LoS. It includes the probability of the event $A_0$, when there are no blockers in the area. Using the law of total probability, we establish
\begin{align}\label{eqn:eA6}
P_{LoS}&=Pr\{A_0\}+\nonumber\\
&+\sum_{i=1}^{\infty}Pr\{A_i\}\prod_{j=1}^{\infty}\left(Pr\{B_0\}+Pr\{B_1|C_0\}\right).
\end{align}

Since the height of a blocker is assumed to be independent from its width, we have $Pr\{B_1|C_0\}=Pr\{B_{1}\}Pr\{C_{0}\}$. Substituting these parameters into (\ref{eqn:eA6}), we arrive at
\begin{align}\label{eqn:eA8}
P_{LoS}&=p_0 + \sum_{i=1}^{\infty} p_{i}\prod_{j=1}^{\infty}\Bigg( \int_{-r_{max}}^{r_{max}}f_R(y) F_R(|y|)dy +\nonumber\\
&+ \int_{-r_{max}}^{r_{max}}\int_{0}^{r}f_R(y) (1 - F_R(|y|) )f_R(x)\times \nonumber\\
&\times F_H\left(\frac{h_{T}-(h_{T} - h_{R})x}{r}\right) \Bigg)dydx.
\end{align}
where $p_{i}$, $i=0,1,\dots$ are the Poisson probabilities.

\section{Numerical results and discussion}\label{sec:numerical}

\subsection{Calibration with simulations}

\begin{figure*}[t]
\centering
\subfigure[\label{fig:rlfailex_a}Scenario modeled with ray-launching]
{\includegraphics[width=.32\textwidth]{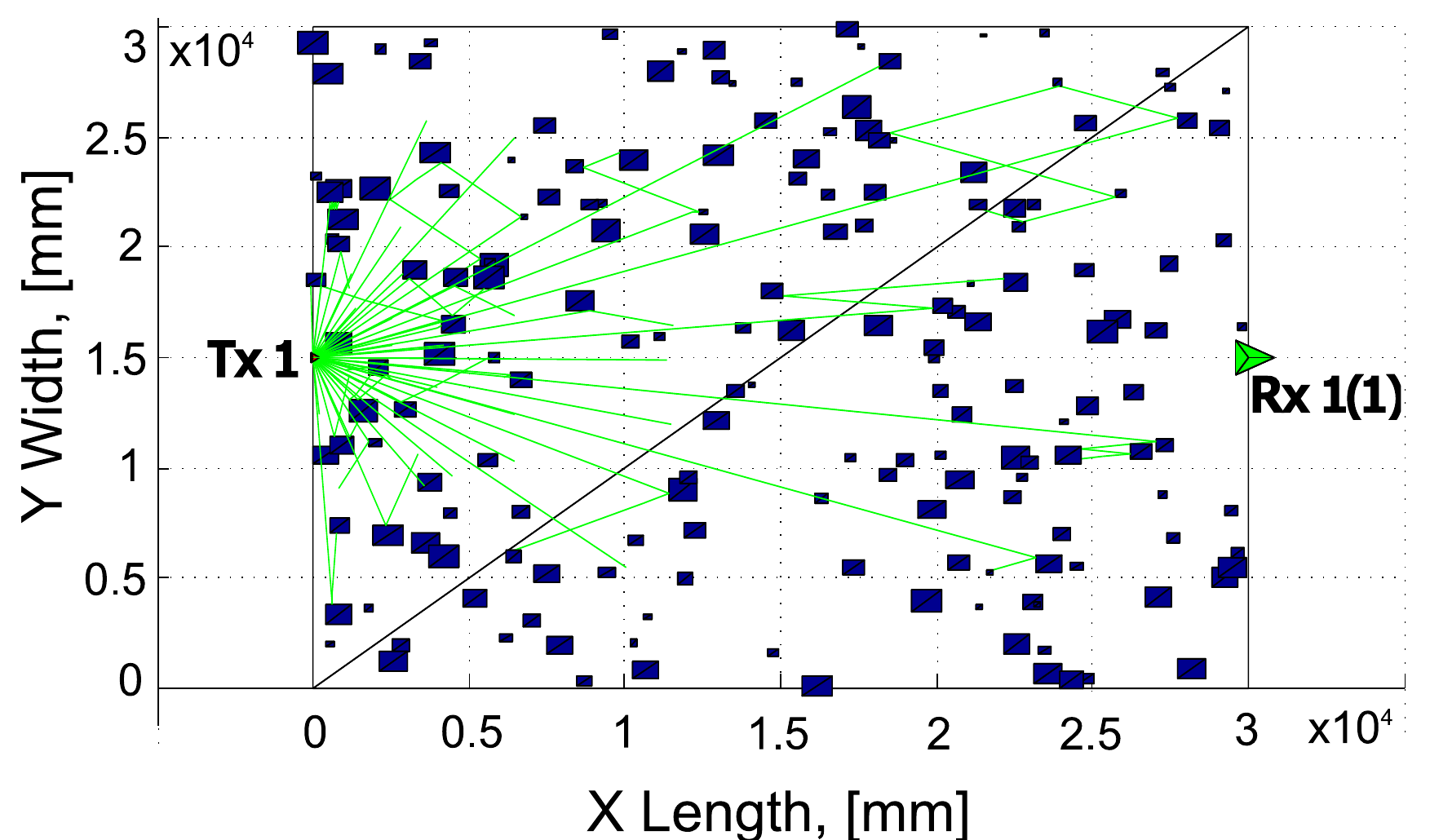}}
\subfigure[\label{fig:rlfailex_b}Path loss calibration with ray-launching]
{\includegraphics[width=.32\textwidth]{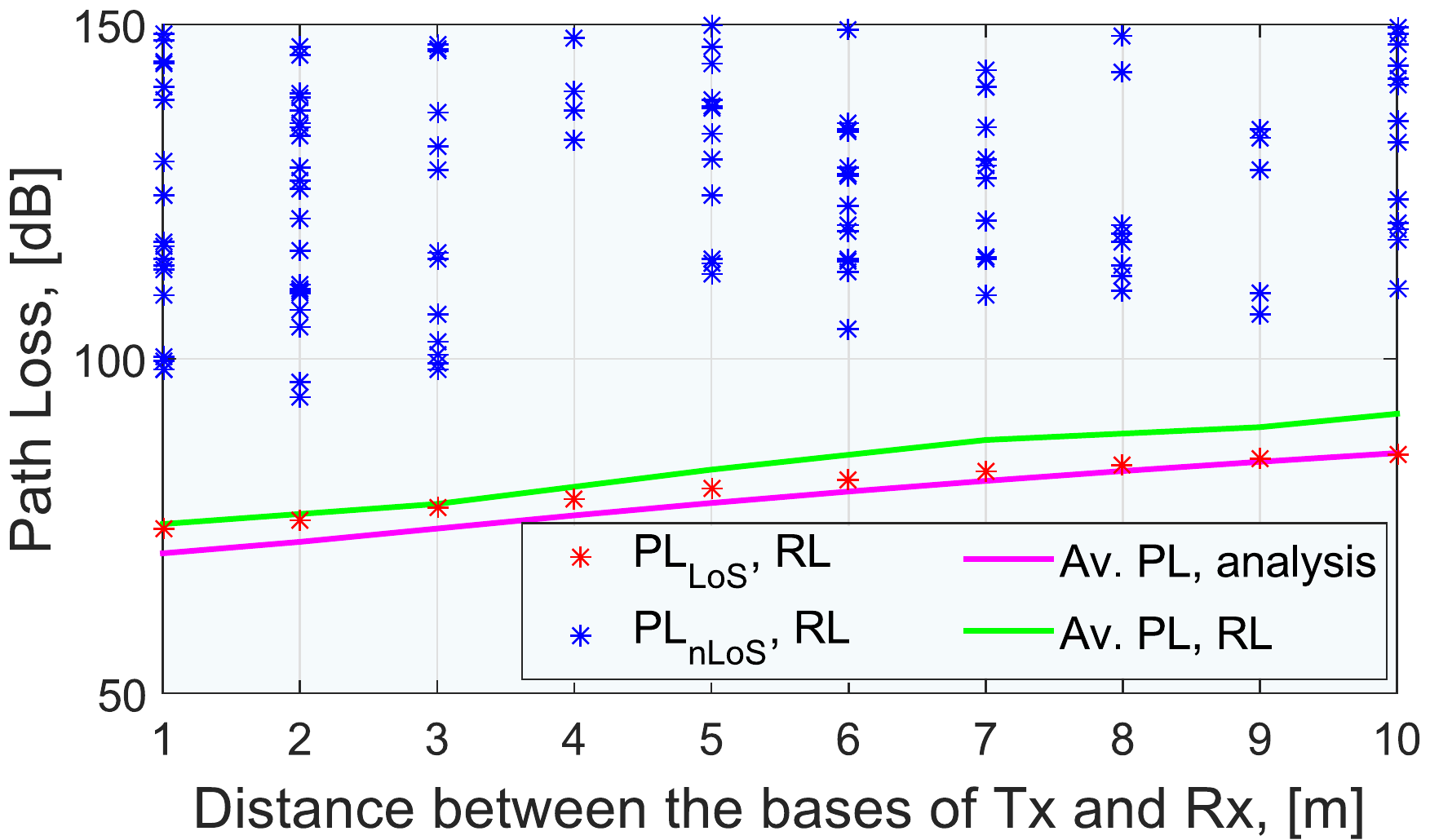}}
\subfigure[\label{fig:rlfailex_c}Blockage probability calibration]
{\includegraphics[width=.32\textwidth]{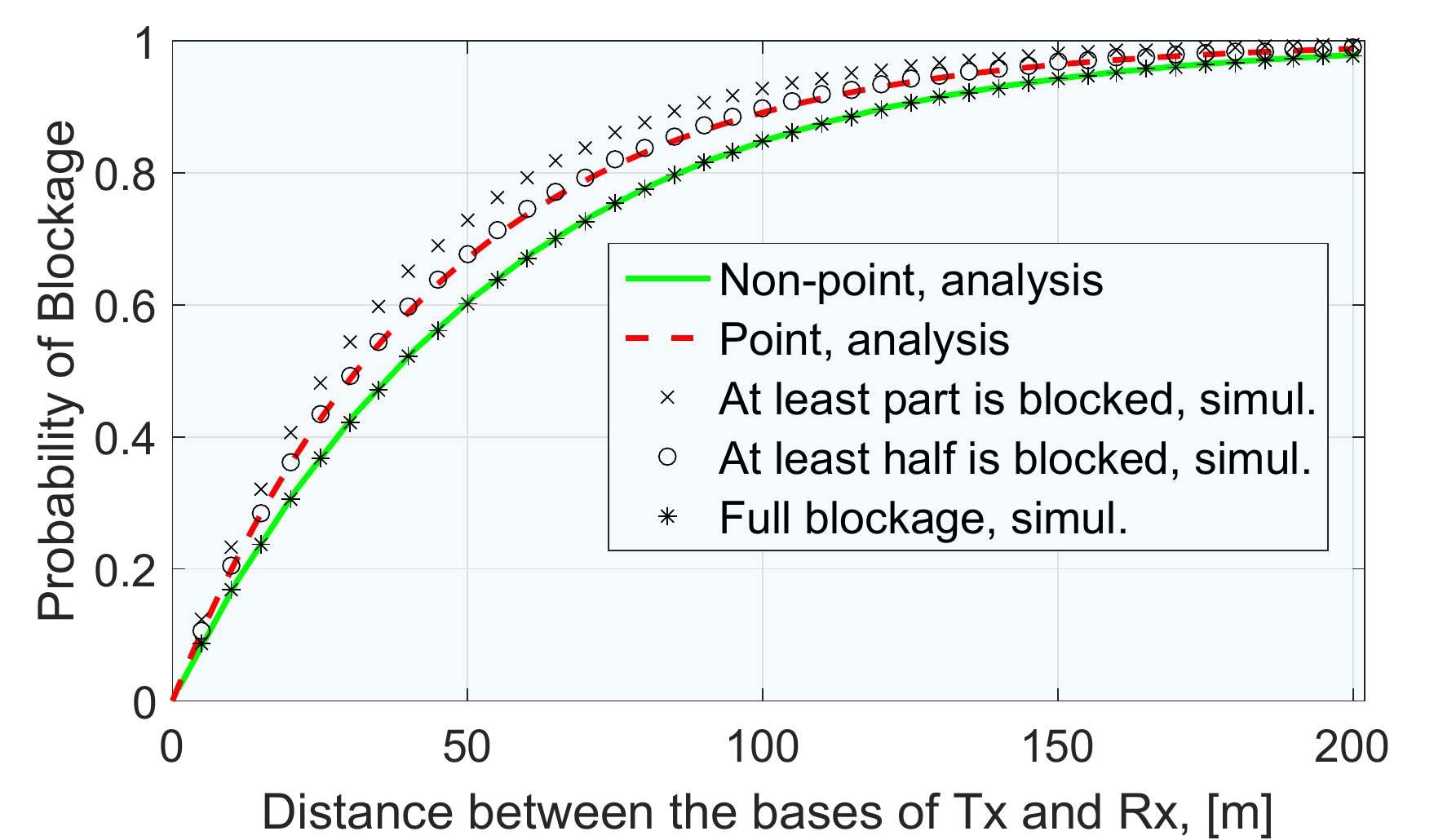}}
\caption{Summary of calibration results.}
\end{figure*}

For the verification of our analytical model, we employed our own mmWave ray-launching simulator, which approximates the propagation of electromagnetic waves with geometric lines. In addition, the tool accounts for detailed reflection effects, which accurately mimic mmWave diffraction, refraction, and scattering. More specifically, the rays intersecting the Rx antenna are considered to be successfully received multi-path components. Based on the information about the ray delays, phase, and received power, any relevant statistical data could be obtained. As a result, our industry-grade ray-launching tool makes it possible to recreate the needed urban environment and obtain results, which are reasonably close to real measurements. A simplified example of a ray-launching based simulation run without scattering and diffraction effects (for simplicity of exposition) is shown in Fig.~\ref{fig:rlfailex_a}.

We continue by calibrating the model with the ray-launching tool based on the path loss comparison. In Fig.~\ref{fig:rlfailex_b}, the results of the simulated LoS and non-LoS path loss are given by red and blue dots, respectively, while the average ray-launching path loss is plotted in green. Finally, the results of our analytical modeling are presented in a form of the average path loss calculated as
\begin{align}\label{eqn:eA777}
L_{e}=P_{LoS}L_{LoS} + (1-P_{LoS})L_{nLoS},
\end{align}
where $P_{LoS}$ is the LoS probability computed according to (\ref{eqn:eA8}), $L_{LoS}$ and $L_{nLoS}$ are the path loss for LoS and non-LoS components, which have been parameterized according to the real measurements in \cite{rappaport}. The comparison between the average ray-launching and the average analytical path loss indicates a marginal difference. However, it should be noted that the results have only been simulated at shorter distances due to excessive computational complexity. In the plot, it is also visible that the LoS path loss values differ from the corresponding non-LoS figures by at least $20$ dB. Based on this observation, we conclude that if the LoS component is not present, there is a high probability of poor mmWave link, similar to complete receiver blockage due to very low SNR.

\begin{table}[b!]\footnotesize
\centering
\caption{Main simulation parameters}
\begin{tabular}{p{5.5cm}p{2.5cm}}
\hline
\textbf{Parameter}&\textbf{Value} \\
\hline
Height of Tx&$4m$\\
Height of Rx&$1.3m$\\
Distance between the bases of Tx and Rx&$30m$\\
Height of a blocker, $N(\mu_{H}, \sigma_{H})$ & $N(1.7m, 0.1m)$\\
Diameter of a blocker, $U(d_{min},d_{max})$ & $U(0.2m,0.8m)$\\
Length of Rx&$0.1m$ \\
Initial intensity of blockers&$0.3$ $blockers/m^{2}$\\
Frequency&$28GHz$\\
\hline
\end{tabular}
\label{tab:sim_parameters}
\end{table}

To facilitate a more detailed comparison, and in addition to ray-launching, we developed a simplified system-level simulator providing the blockage probability as a function of the model parameters, which are given in Table \ref{tab:sim_parameters}. First, in a square area of interest with the dimensions of ten times greater than the distance between the Tx and Rx, we generated the blockers according to the modified Matern hard-core process with the parameters $\lambda_{I}$ and $U(0,d_{\max}/2)$ following the algorithm described in \cite{2}. Here, the Tx is assumed to be a point and is placed in the center, while the Rx is a segment with the length of $l$ positioned at the distance $r$ from Tx, such that the line connecting the center of Rx and Tx is a normal. Both Tx and Rx are associated with heights $h_{T}$ and $h_{R}$, respectively. We further divide the segment $l$ into $N$ parts. In these settings, (i) the event when the Rx is fully blocked occurs when all of the lines are blocked; (ii) the event that at least half of the Rx is blocked corresponds to having at least half of the lines blocked; and (iii) the event that at least part of the Rx is blocked occurs when at least one line is blocked.

The comparison of the results obtained with simulations and the proposed mathematical model for point and interval Rx of size 10cm is shown in Fig.~\ref{fig:rlfailex_c}. First, notice that the maximum absolute difference is attained at the average distances of around $40-100$ meters. For short and long distances, the deviation is marginal. Nevertheless, the maximum difference is always less than $0.1$ for a reasonably large Rx of 10cm. The slight deviations of the analytical results from simulations are explained by the fact that we replaced the actual Matern hard-core process by a Poisson process with the same intensity.

\subsection{Understanding analytical results}



\begin{figure}[!h]
\centering
\includegraphics[width=\columnwidth]{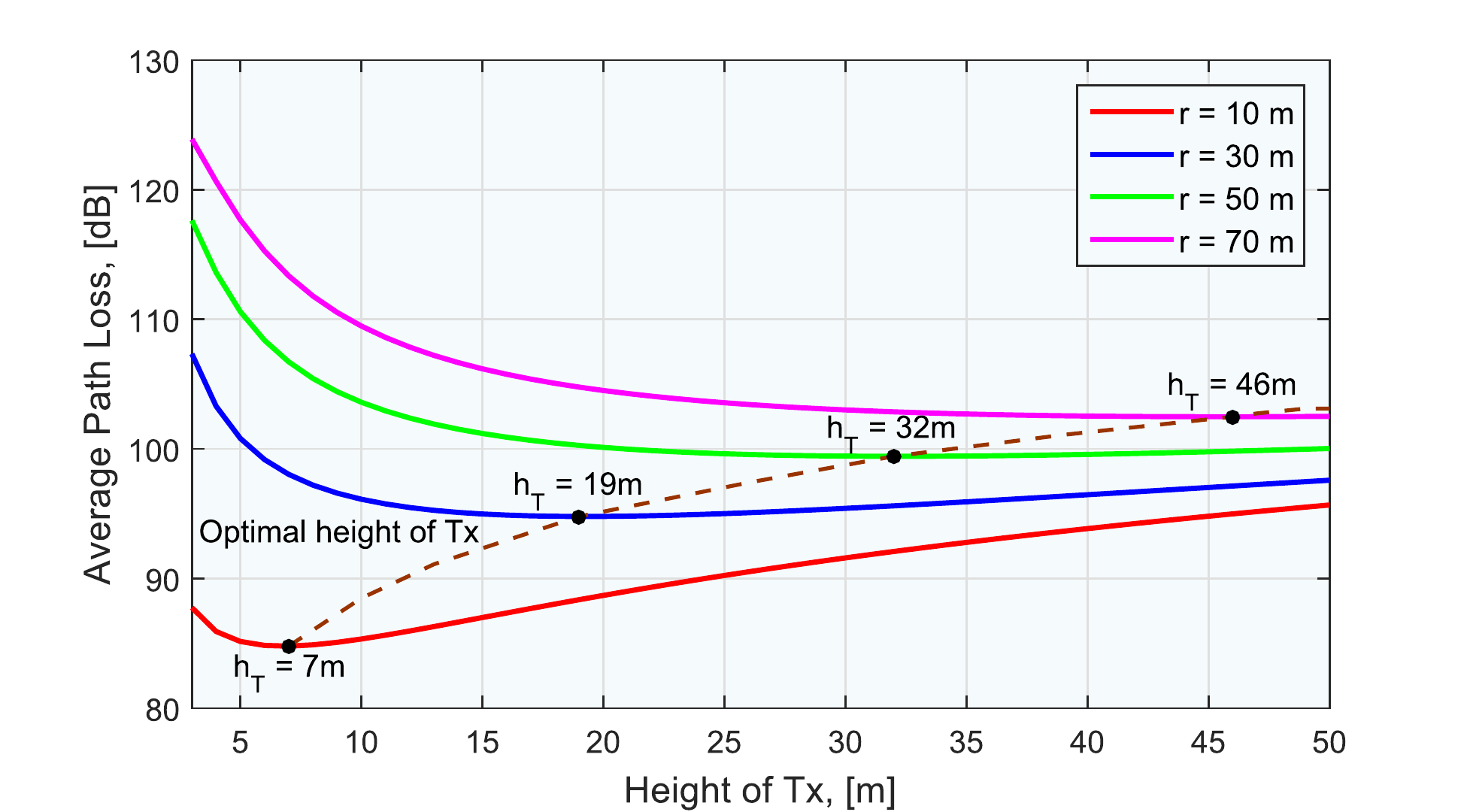}
\caption{Average path loss and optimal $h_{T}$ for different $r$.}
\label{fig:Plot3}
\end{figure}

First, Fig.~\ref{fig:Plot3} shows the results for the average path loss ($L_{e}$) as a function of the Tx height calculated according to (\ref{eqn:eA777}). As one may observe, for each separation distance between the bases of Tx and Rx, there always exists an optimal height of the Tx, where the average path loss takes the minimum value.

\begin{figure}[h!!]
\centering
\includegraphics[width=\columnwidth]{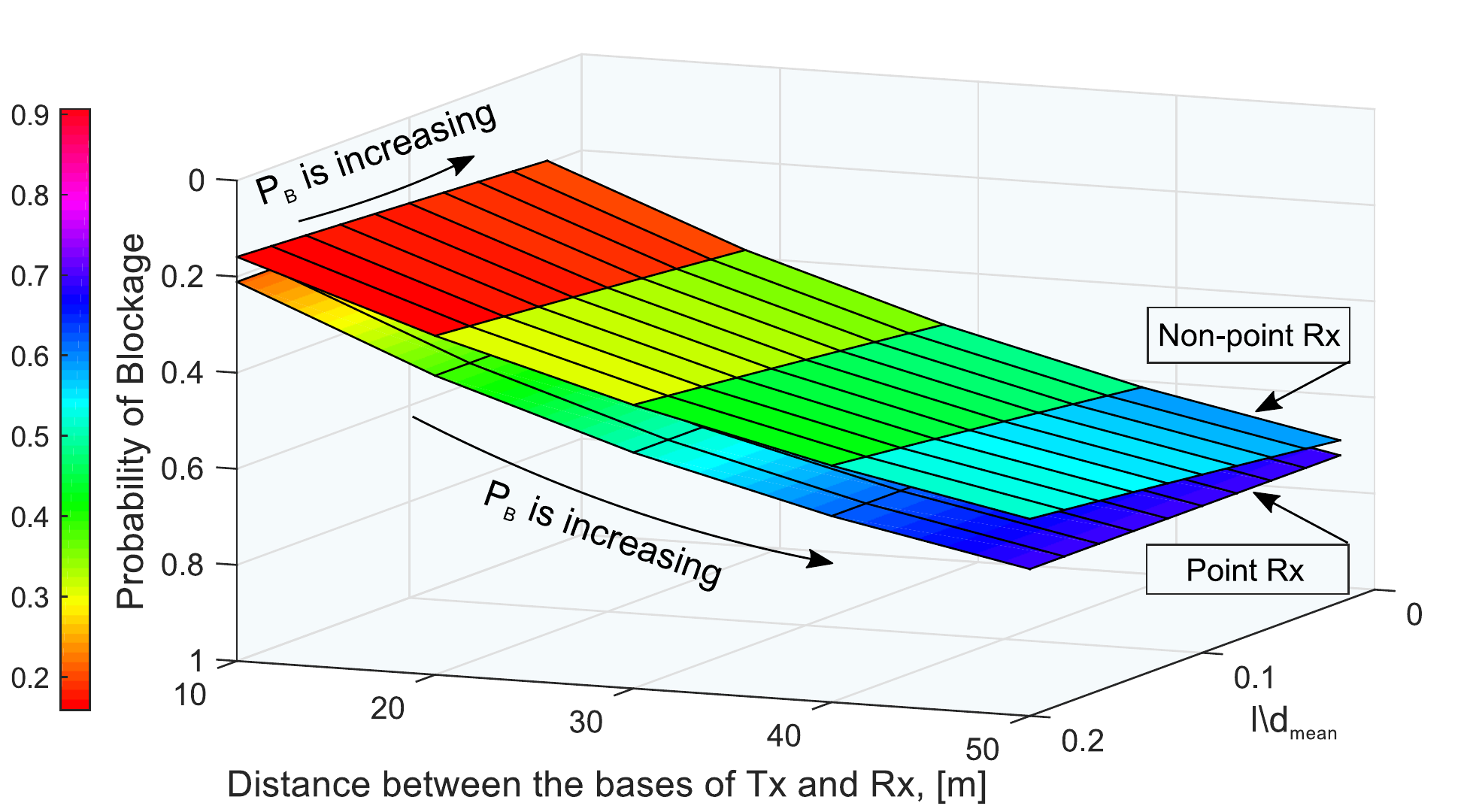}
\caption{Blockage probability as a function of $r$ and $l/d_{mean}$.}
\label{fig:Plot2}
\end{figure}

To characterize the effect of non-infinitesimal Rx, Fig.~\ref{fig:Plot2} demonstrates the blockage probability for the interval Rx across different ratios of the Rx size and the mean blocker diameter $l/d_{mean}$, as well as the distance between the bases of Tx and Rx $r$. We notice that for any value of $r$ and $l/d_{mean}$, the blockage probability for the point Rx is slightly higher than that for the interval Rx. The increase in $l/d_{mean}$, however, decreases the blockage probability.

The height of the Tx, $h_{T}$, is one of the most important parameters available for system designers. Intuitively, the higher the height is, the smaller the probability of the LoS blockage should be. However, the final effect is expected to heavily depend on the Tx-Rx separation distance.

\begin{figure}[!h]
\centering
\includegraphics[width=\columnwidth]{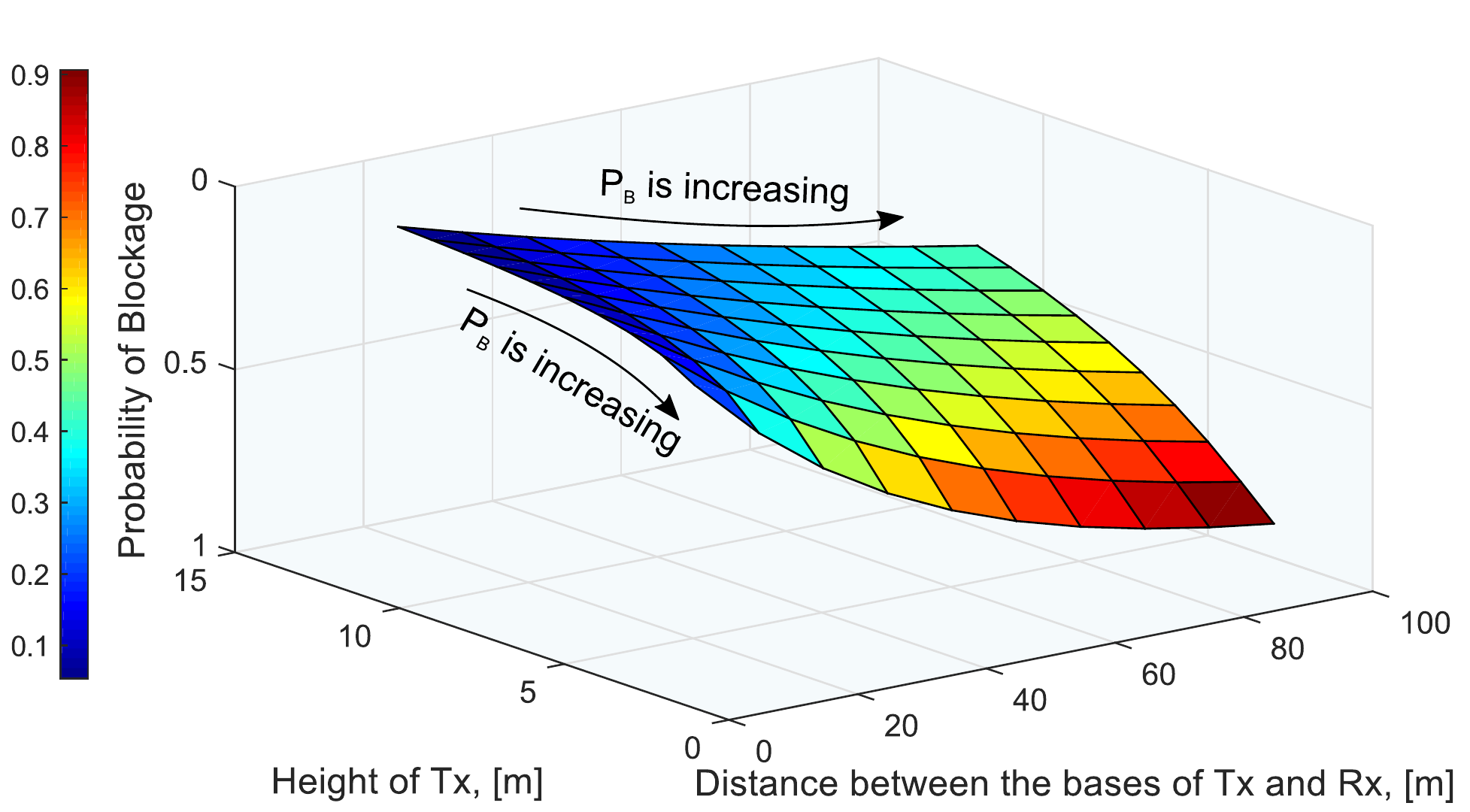}
\caption{Blockage probability for different $h_{T}$ and $r$.}
\label{fig:Plot4}
\end{figure}

The impact of the Tx height and distance between the Tx and Rx for a point Rx is demonstrated in Fig.~\ref{fig:Plot4}. As we expected, the probability of blockage reduces exponentially as $h_{T}$ increases. The effect of the Tx-Rx distance is however inverse -- longer separation distances lead to higher blockage probability. For large $h_{T}$, the increase is linear, whereas for smaller $h_{T}$ it is exponential. Also, notice that starting from a certain Tx-Rx separation distance, the blockage probability remains nearly the same. It is explained by the fact that at such distances, and for a given human user density $\lambda_{I}=0.3$, the probability to ''meet'' a blocker along the LoS path is high. 

\section{Conclusion}\label{sec:conclusions}


The presence of the LoS signal path is crucial in determining the ultimate performance of the 5G-grade mmWave wireless communications, especially in highly-crowded outdoor environments. Using the ray-launching simulator, we first demonstrated that the difference between the LoS and non-LoS path loss values is at least $20$ dB. Moreover, due to shorter ranges of mmWave cellular technology, smaller obstacles will affect the corresponding quality of signal propagation. All of the above leads to the need of having a comprehensive analytical methodology for determining the presence of the LoS link. However, existing approaches, such as 3GPP urban outdoor micro-cellular model, do not deliver the desired detalization. This substantiates the importance of our conducted LoS-centric analysis.

To this end, we developed a novel framework for capturing the effects of the most important mmWave system parameters on the probability of blockage for both the infinitesimal receiver and the receiver of a fixed non-zero length. As a result, it was shown that an increase in the ratio between the length of the Rx and the blocker sizes leads to a closer match between the two. Moreover, we demonstrated the existence of the optimal Tx height, which is an important learning for system designers. The studied parameters of interest included the distance between the bases of the Tx and Rx and their heights, as well as the density of blockers on the landscape and their random dimensions.

Finally, also with the help of an advanced ray-launching tool, it was confirmed that our analysis offers accurate results to model human-body blockage in urban mmWave deployments.

\section{Acknowledgements}\label{sec:acknowledgements}
 The authors would like to thank Dmitrii Solomitckii, Tampere University of Technology, for helping with ray-launching simulations.  This work is supported by Intel Corporation.


\begin{thebibliography}{99}


\bibitem{Andrews2} J.G. Andrews, S. Buzzi, C. Wan, S.V. Hanly, A. Lozano, A.C.K. Soong, J.C.Zhang, "What will 5G be?," \textit{IEEE Journal on Selected Areas in Communications}, vol. 32, no. 6, pp. 1065--1082, June 2014.
\bibitem{propagation} C.R. Anderson, T.S. Rappaport, "In-building wideband partition loss measurements at 2.5 and 60 GHz," \textit{IEEE Transactions on Wireless Communications}, vol. 3, no. 3, pp. 922--928, May 2004.
\bibitem{standard} 3GPP TR 36.814 V9.0.0  “Evolved Universal Terrestrial Radio Access (E-UTRA); Further advancements for E-UTRA physical layer aspects (Release 9)”, version 9.0.0, March 2010.
\bibitem{heath2} T. Bai, R. Vaze, R. W. Heath, Jr., "Analysis of Blockage Effects on Urban Cellular Networks," \textit{IEEE Transactions on Wireless Communications}, pp. 5070--83, September 2014.
\bibitem{SinJSAC14} S. Singh, M. N. Kulkarni, A. Ghosh, J. G. Andrews, "Tractable Model for Rate in Self-Backhauled Millimeter Wave Cellular Networks," \textit{IEEE Journal on Selected Areas in Communications}, vol. 33, no. 10, pp. 2196-2211, October 2015.
\bibitem{rappaport} M. R. Akdeniz, Y. Liu, M. K. Samimi, S. Sun, S. Rangan, T. S. Rappaport, E. Erkip, "Millimeter Wave Channel Modeling and Cellular Capacity Evaluation," \textit{IEEE Journal on Selected Areas in Communications}, vol. 32, no. 6, pp. 1164--79, June 2014.
\bibitem{andrews} M. N. Kulkarni, S. Singh, J. G. Andrews, "Coverage and rate trends in dense urban mmWave cellular networks," \textit{IEEE Global Communications Conference (GLOBECOM)}, pp. 3809--14, December 2014.
\bibitem{cylinders} M. Jacob, S. Priebe, T. Kurner, M. Peter, M. Wisotzki, R. Felbecker, W. Keusgen, "Fundamental analyses of 60 GHz human blockage," \textit{7th European Conference on Antennas and Propagation (EuCAP)}, pp. 117--121, April 2013.
\bibitem{height} C. L. Ogden, C. D. Fryar, M. D. Carroll, K. M. Flegal, "Mean Body Weight, Height, and Body Mass Index, United States 1960--2002," \textit{Centers for Disease control and prevention},  no. 347, October 2004.
\bibitem{Andrews1} J. Andrews, K. Ganti, M. Haenggi, N. Jindal, S. Weber, "A primer on spatial modeling and analysis in wireless networks," \textit{IEEE Communications Magazine}, vol. 48, no. 11, pp. 156--163, November 2010.
\bibitem{2} S. N. Chiu, D. Stoyan, W. S. Kendall, J. Mecke, "Stochastic Geometry and its applications," \textit{Wiley}, 3d ed., 2013.
\bibitem{ross} S. Ross, "Introduction to probability models," \textit{Academic Press}, 10th ed., 2009.
\bibitem{cox_Miller} D. R. Cox, H.D. Miller, "The theory of stochastic processes," \textit{Wiley}, 1965.
\bibitem{cox} D. R. Cox, "Renewal theory," \textit{Methuen and Co ltd}, 1970.\\


\end{thebibliography}
\end{document}